\documentclass[aip,jap,reprint]{revtex4-1}

\usepackage{graphicx}
\usepackage{dcolumn}
\usepackage{bm}
\usepackage{amssymb}
\usepackage{amsmath}
\usepackage{wrapfig}
\begin{document}

\title{An analog magnon adder for all-magnonic neurons}

\author{T. Br\"acher}
\affiliation{Fachbereich Physik and Forschungszentrum OPTIMAS, Technische Universit\"at
Kaiserslautern, D-67663 Kaiserslautern, Germany}
\author{P. Pirro}
\affiliation{Fachbereich Physik and Forschungszentrum OPTIMAS, Technische Universit\"at
Kaiserslautern, D-67663 Kaiserslautern, Germany}

\date{\today}

\begin{abstract}
Spin-waves are excellent data carriers with a perspective use in neuronal networks: Their lifetime gives the spin-wave system an intrinsic memory, they feature strong nonlinearity, and they can be guided and steered through extended magnonic networks. In this work, we present a magnon adder that integrates over incoming spin-wave pulses in an analog fashion. Such an adder is a linear prequel to a magnonic neuron, which would integrate over the incoming pulses until a certain nonlinearity is reached. In this work, the adder is realized by a resonator in combination with a parametric amplifier which is just compensating the resonator losses.
\end{abstract}


\maketitle

\section{Introduction}
In certain tasks like pattern recognition, the brain outperforms conventional CMOS-based computing schemes by far in terms of power consumption. Consequently, neuromorphic computing approaches aim to mimic the functionality of neurons in a network to boost computing efficiency\cite{Neuro1,Neuro2,Neuro3, Neuro4, Neuro5}. In the brain, stimuli are conveyed by short wave packets from one neuron to another, where they lead to stimulation which adds up and then, ultimately, triggers a nonlinear response. Thus, it is natural to consider waves as data carriers for bio-inspired computing and artificial neuronal networks. Certain key components need to be accessible by the used kind of waves: It should be possible to convey them through extended networks as well as to store the information carried by the waves for a certain time, so that stimuli can add up. In addition, the waves should exhibit nonlinear dynamics in order to mimic the threshold characteristics of a neuron. Among the possible waves that one can consider, spin waves, the collective excitation of magnetic solids, are a highly attractive candidate\cite{Magnon-Spintronics, Neusser, Lenk-2011-1, Magnonics-Krug, Mag-Road}: The dynamics of spin-waves and their quanta, magnons, are governed by a nonlinear equation of motion\cite{Stancil, Melkov}, providing easy access to nonlinearity\cite{Pirro-2014-PRL, Bauer-2015-1, Slavin-1991-JAP}. They can be guided through reprogrammable networks by using spintronics and nonlinear effects and their finite lifetime provides an intrinsic memory to the spin-wave system. In addition, their excitation energy is very low and their nanometric wavelengths at frequencies in the GHz and THz range promise a scalable and power-efficient platform for neuromorphic computing.

\begin{figure}[h]
	  \begin{center}
    \scalebox{1}{\includegraphics[width=8 cm, clip]{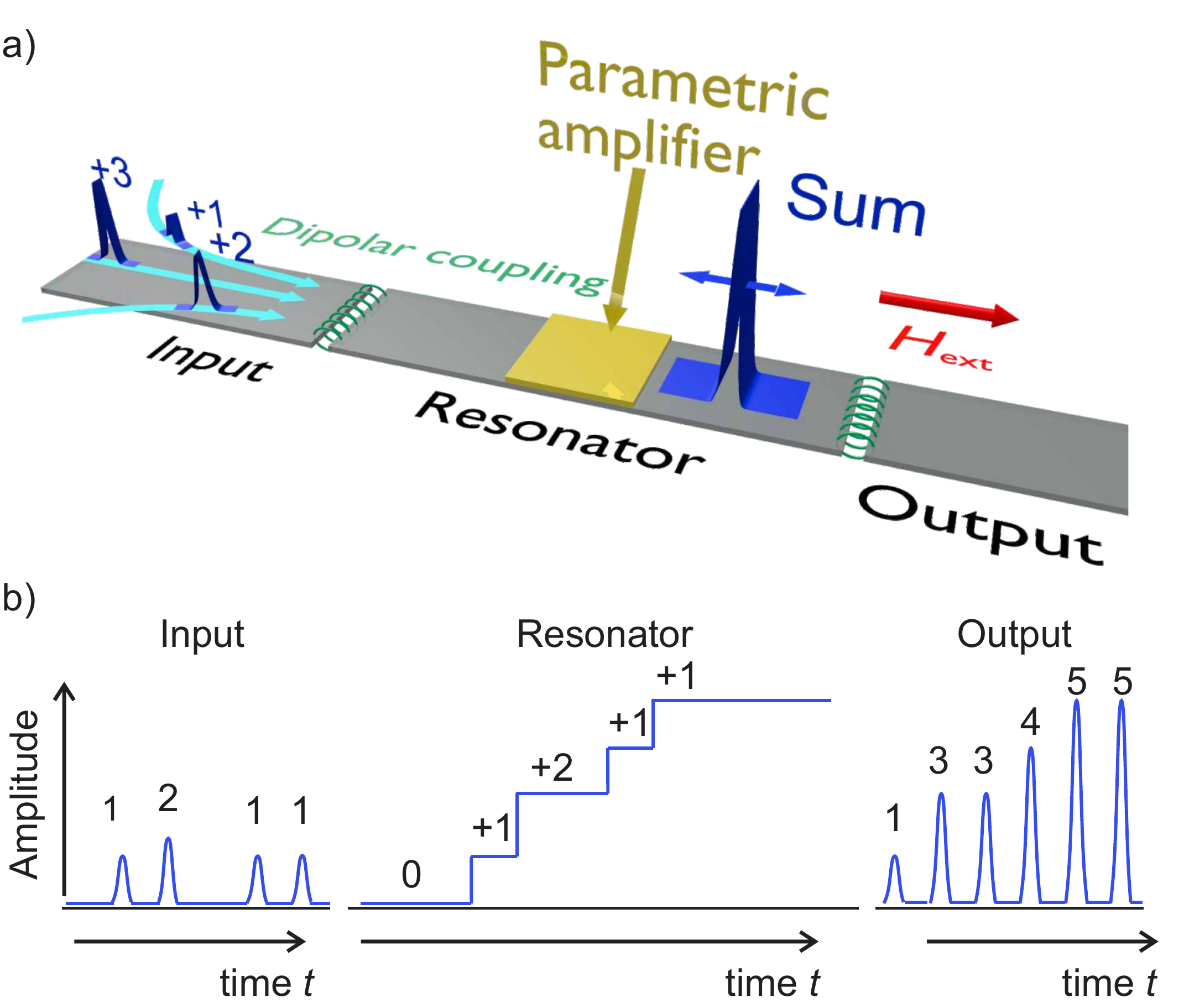}}
    \end{center}
	  \caption{\label{Fig1}a) Sketch of the magnon adder consisting of an input, a resonator and an output, as well as a parametric amplifier to compensate the losses in the resonator. b) Sketch of the operation principle of the amplifier: Subsequent pulses with amplitude $A_p(n)$ (indicated by the numbers) enter the resonator, which integrates over their amplitude. Periodically, a pulse leaves the resonator at the output. The value stored in the resonator and the value in the output are equal to the sum $S$ of the input amplitudes.}
\end{figure}

In this work, we employ micromagnetic simulations to demonstrate an analog magnon adder, which can be regarded as a pre-step to a magnon based neuron. The adder, which is sketched in Fig.~\ref{Fig1}~(a), consists of two building blocks: A leaky spin-wave resonator and a parametric amplifier\cite{Review-PP, Braecher-2014-APL1, Lvov, PtI}. Spin waves can enter the resonator by dipolar coupling to the input\cite{Tunnel}. Within it, their amplitude is added to or subtracted from the amplitude of the already accumulated amplitudes, as is sketched in Fig.~\ref{Fig1}~(b). This process is, in principle, equivalent to the arrival of excitation pulses in the axon, where the neuron integrates over the incoming stimuli until a critical stimulus is reached. In our scheme, the parametric amplifier acts to counteract the spin-wave losses that arise from propagation through the resonator and the leakage to the input and output of the resonator. We show that by working at the point of loss compensation, the adder can add and subtract the spin-wave amplitudes over a large range and enables to store the sum of these calculations in the resonator.

\section{Layout and working principle}
To demonstrate the magnonic adder, we perform micromagnetic simulations using MuMax3\cite{Mumax}. For our simulations, we assume the material parameters of Yttrium Iron Garnet (YIG)\cite{YIGMagnonics,YIGDubs}, a widely used material in magnonics\cite{Saga,Magnon-Spintronics}: Saturation magnetization $M_\mathrm{s} = 140\,\mathrm{kA}\,\mathrm{m}^{-1}$, exchange constant $A_\mathrm{ex} = 3.5\,\mathrm{pJ}\,\mathrm{m}^{-1}$, and Gilbert damping parameter $\alpha = 0.0002$, which represents the damping of the spin-waves mainly into the phonon system. The geometry we study consists of three $w = 0.5\,\mu\mathrm{m}$ wide and $40\,\mathrm{nm}$ thick rectangular YIG waveguides in a row (see Fig.~\ref{Fig1}~(a)). The length of the central waveguide, which acts as the resonator, is $L = 20\,\mu\mathrm{m}$. The waveguide to the left of the resonator acts as input, where spin-waves are excited by a source creating a local magnetic field. In a magnonic network, this input could be connected to an arbitrary number of other waveguides acting as individual inputs. The waveguide on the right of the resonator acts as output, which, again could be reconnected in a network. In our simulations, input and output are $10\,\mu\mathrm{m}$ long and separated from the resonator by $g = 75\,\mathrm{nm}$ wide gaps. Spin-waves can tunnel through this gap\cite{Tunnel}, which constitutes the coupling channel from the resonator to the input and output, respectively. In the present simulation, about $0.1\%$ of the spin-wave amplitude is tunneling through the gap. Towards their outer edges, the damping in the input and output is increased exponentially, to mimic the transport of spin-waves out into the network that would take place in a real extended system. 

\begin{figure}[h]
	  \begin{center}
    \scalebox{1}{\includegraphics[width=8 cm, clip]{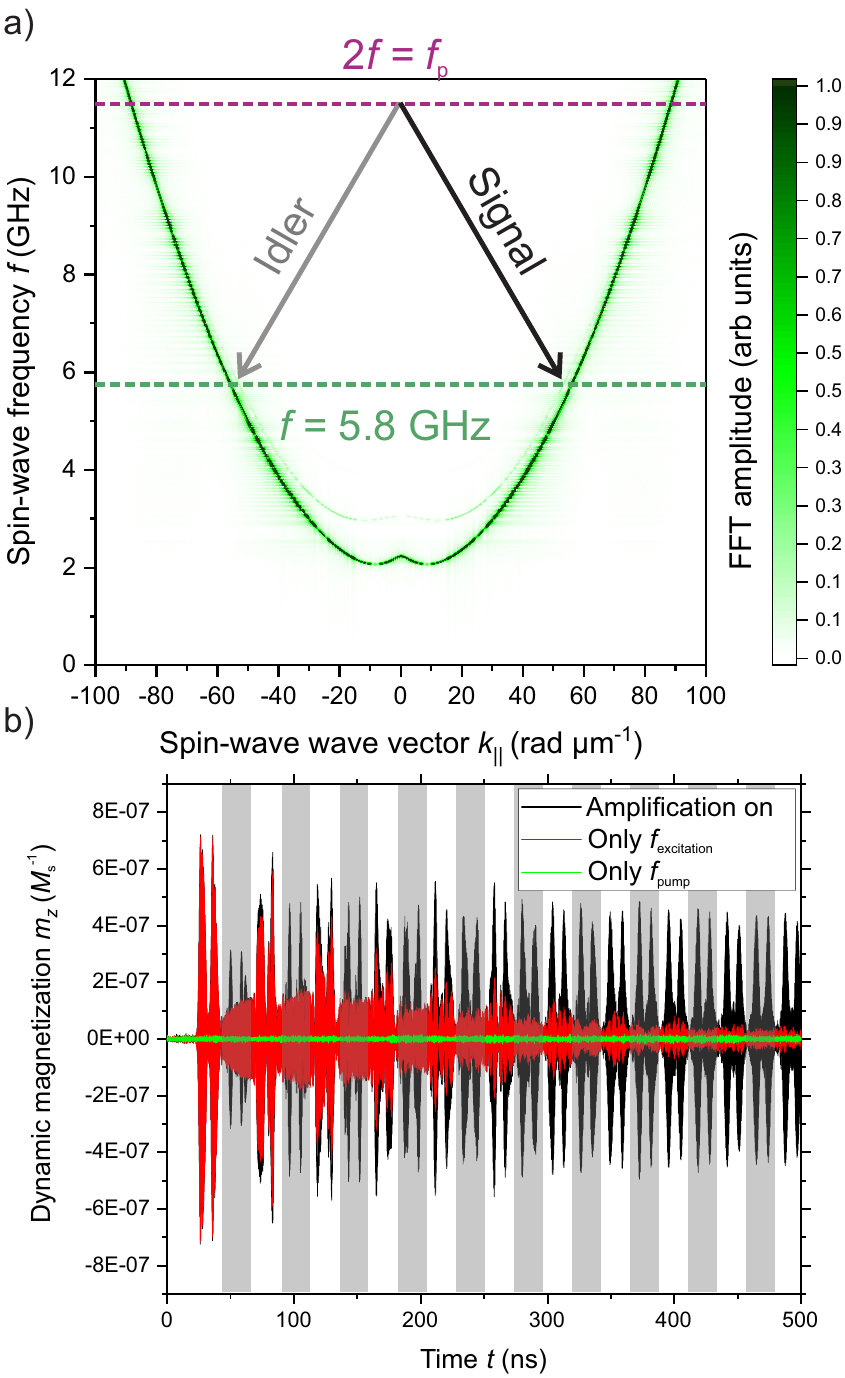}}
    \end{center}
	  \caption{\label{Fig2}a) Simulated spin-wave dispersion relation at a field of $\mu_0H_\mathrm{ext} = 20\,\mathrm{mT}$ applied along the resonator long axis and illustration of the parallel pumping process. b) Dynamic magnetization $6\,\mu\mathrm{m}$ away from the center of the resonator as a function of time. Red: Only excitation of a single spin-wave pulse with amplitude '1' in the input. Green: Only periodic application of amplification pulses, no stimulus at the input. Black: Periodic application of an amplification pulse twice per roundtrip with an input stimulus of one pulse with amplitude '1'. The gray shaded areas indicate the position of the idler waves.}
\end{figure}

Figure~\ref{Fig2}~(a) shows the simulated spin-wave dispersion\cite{Braecher-2017-PRB} of the fundamental mode in a color-coded scale. The external field of $\mu_0H_\mathrm{ext}=20\,\mathrm{mT}$ is applied along the long axis of the resonator. The excitation frequency $f = 5.8\,\mathrm{GHz}$ corresponds to the excitation of spin-waves with a wave-vector of $k_{||} = 56\,\mathrm{rad}{\mu}m^{-1}$ (i.e., wavelength $\lambda \approx 112\,\mathrm{nm}$) and the periodic excitation source is matched to excite this wave-vector resonantly. From the simulated spin-wave dispersion, a group velocity of $v_\mathrm{g} = (0.88\pm0.05)\,\mu\mathrm{m}\,\mathrm{ns}^{-1}$ can be extracted. This corresponds to a roundtrip time of $\Delta t = 2\cdot L/v_\mathrm{g} \approx 46\,\mathrm{ns}$ through the resonator. During one trip, the spin-wave amplitude $n$ decays exponentially following $A_\mathrm{p}(t) = A_\mathrm{p}(0) \cdot \mathrm{exp}(-t/\tau)$ with their lifetime $\tau$. From this, it can be inferred that during one pass lasting $\Delta t$, the relative amplitude change is:
\begin{align}
\label{Eq1}
\frac{A_\mathrm{p}(t+\Delta t)}{A_\mathrm{p}(t)} = \mathrm{e}^{\frac{-\Delta t}{\tau}}.
\end{align}
As mentioned above, the dipolar coupling between the resonators is very weak and only a small fraction of $0.1\%$ of the spin-wave amplitude is actually coupled from the resonator to the input and output, respectively. Consequently, the losses of the resonator are dominated by the propagation loss.

In order to counteract these losses, we employ parametric amplification, also known as parallel pumping\cite{Review-PP, Braecher-2014-APL1, Lvov, PtI}. In this technique, the system is pumped at the frequency $f_\mathrm{p}$ which equals twice the resonance frequency $f$. One possible driving force are Oersted fields\cite{Review-PP,PtI} $\mu_0 h_\mathrm{p}$, where microwave photons split into pairs of magnons as indicated in Fig.~\ref{Fig2}~(a). Here, we consider this mechanism, but also other, more energy efficient realizations like the use of electric fields have been proposed\cite{Verba1, Verba2}. In the simplest case of adiabatic parametric amplification\cite{Review-PP}, the pumping at $2f$ leads to the formation of wave pairs at $f$ with wave-vector $\pm k_{||}$ in order to conserve momentum, as is sketched in Fig.~\ref{Fig2}~(a). Parallel pumping counteracts the damping losses with two key features\cite{Review-PP}: 1. It only couples to already existing waves and, in the absence of nonlinear saturation, leads to an exponential increase of the spin-wave amplitude following $A_\mathrm{p}(t) = A_\mathrm{p}(0) \cdot \mathrm{exp}((V \mu_0 \cdot h_p-\tau^{-1})t)$ if the energy per unit time $V \cdot \mu_0 h_\mathrm{p}$ inserted into the spin-wave system exceeds the losses given by $\tau^{-1}$. Here, $V$ constitutes the coupling parameter of the given spin-wave mode at $(f,k_{||}$). 2. Parallel pumping conserves the phase of the incident spin-waves. This is important to profit from the phase of the spin-wave in encoding which is, for instance, vital to be able to perform subtraction in the presented magnon adder. In the simulated structure, the local parametric amplifier exhibits an extent of $1\mu\mathrm{m}$ along the resonator and it is situated in the center of the resonator. For simplicity, we only take into account the parallel component of the microwave field created by a stripline\cite{APL-PPBV, PtI}. 

Similar to the operation in the brain, we assume that incident information is carried by pulses. As sketched in Fig.~\ref{Fig1}~(a), these pulses can arrive at the amplifier with different amplitudes and at different times. They exhibit a fixed duration of $5\,\mathrm{ns}$ and delayed pulses are send to the input at times which are integer multiples of the roundtrip time $\Delta t$. The amplification is also pulsed: $t_\mathrm{p}=5\,\mathrm{ns}$ long pumping pulses are applied whenever the spin-wave pulse in the resonator passes the amplifier, i.e., twice per roundtrip. When the net increase of the spin-wave amplitude by the pumping is equivalent to the losses, this leads to the formation of a pair of signal and idler spin-waves running back and forth in the resonator. 
 The general act of the parametric amplification is shown in Fig.~\ref{Fig2}~(b) for one single input pulse of amplitude $'1'$. In our simulations, this amplitude was arbitrarily chosen to correspond to an external excitation with a local field amplitude of $65\,\mu\mathrm{T}$ in the input. The diagram shows the out-of-plane dynamic magnetization component $m_z$ as a function of time at a point $6\,\mu\mathrm{m}$ away from the resonator center. The red curve shows the amplitude if no pumping field is applied - the spin-waves pass the position where $m_z$ is recorded for the first time at $t = t_0 \approx 27\,\mathrm{ns}$. They are reflected at the end of the resonator and pass the measurement position again at $t \approx 36\,\mathrm{ns}$. Then they pass a roundtrip through the resonator and arrive again at $t = t_0+\Delta t \approx 73\,\mathrm{ns}$ and so forth. The damping of the waves can be clearly seen and it amounts to about $25\%$ per roundtrip. In contrast, the black curve shows the time evolution of the spin-wave amplitude if the parametric amplification is switched on and is just strong enough to compensate the losses during one roundtrip. Now, the idler pulses are created, which give rise to additional pulses highlighted by the gray shaded areas. After the idler is build up and after some initial fluctuations, the quasi steady-state is reached and the pulses run back and forth with constant amplitude. For completeness, the green curve shows the dynamic magnetization if only the pumping pulses are applied, showing that for the presented parameters, noise creation by parametric generation is negligible\cite{Review-PP, Braecher-2011-APL}.

\section{Working point of the magnonic adder}
\begin{figure}[b]
	  \begin{center}
    \scalebox{1}{\includegraphics[width=8 cm, clip]{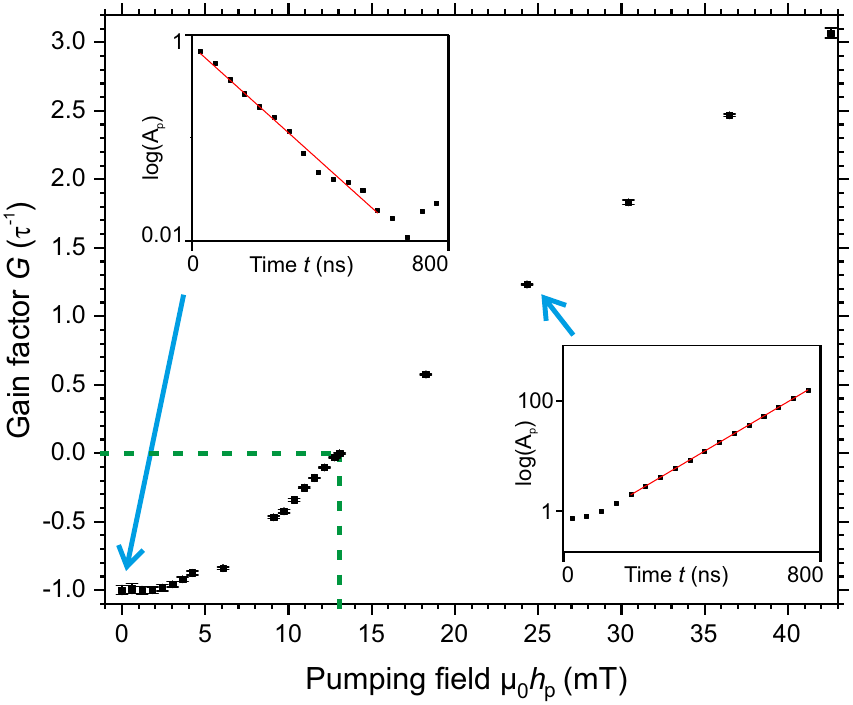}}
    \end{center}
	  \caption{\label{Fig3}Gain factor as a function of the applied pumping field $\mu_0 h_\mathrm{p}$. The insets show the amplitude of the spin-wave pulses $A_\mathrm{p}$ as a function of time for the case of $\mu_0 h_\mathrm{p} = 0$ ($G = -1$, upper inset) and $\mu_0 h_\mathrm{p}=24.3\,\mathrm{mT}$ ($G = +1.2$, lower inset) on a semi-logarithmic scale. The adder is operated at the damping compensation point $G = 0$, marked by the dashed lines.}
\end{figure}
In the following, we want to elaborate the impact of the parametric amplification in more detail, since it plays a crucial role for the operation of the resonator as an adder or as a nonlinear device. The absolute gain per roundtrip is determined by the strength of the pumping field $\mu_0 h_\mathrm{p}$. For half a roundtrip and assuming that $t = 0$ is the point in time when the pulse enters the amplifier, we can modify Eq.~\ref{Eq1} to:
\begin{align}
\label{Eq2}
\begin{split}
A_\mathrm{p}\left(\frac{\Delta t}{2}\right) &= A_\mathrm{p}(0) \cdot \mathrm{e}^{\left(V \mu_0 h_\mathrm{p} - \tau^{-1}\right)\Delta t_\mathrm{p}}  \cdot \mathrm{e}^{- \tau^{-1}\left(\frac{\Delta t}{2}-\Delta t_\mathrm{p}\right)}\\
&= A_\mathrm{p}(0) \cdot \mathrm{e}^{V \mu_0 h_\mathrm{p} \Delta t_\mathrm{p} - \tau^{-1} \frac{\Delta t}{2}}\\
&= A_\mathrm{p}(0) \cdot \mathrm{e}^{0.5\cdot G'(h_\mathrm{p})},  
\end{split}
\end{align}
with the gain $G'(h_\mathrm{p}) = 2 V \mu_0 h_\mathrm{p} \Delta t_\mathrm{p} - \tau^{-1} \Delta t$ per roundtrip. In the following, we will consider a normalized gain factor $G = G'(h_\mathrm{p}) / G'(0) = G'(h_\mathrm{p}) /(\tau^{-1} \cdot \Delta t)$, which is $-1$ in the absence of parametric amplification, $0$ when the parallel pumping is just compensating the losses, and which takes positive values if more energy is inserted per roundtrip than is lost by dissipation. Figure~\ref{Fig3} shows the gain factor $G$ as a function of the applied pumping field, which has been extracted from a linear fit to $\mathrm{ln}(m_z(t)) \propto \mathrm{ln}(A_\mathrm{p}(t))$ as is exemplarily shown in the insets for $\mu_0 h_\mathrm{p} = 0$ ($G=-1$), corresponding to the intrinsic spin-wave decay with the lifetime $\tau = 155\,\mathrm{ns}$, and $\mu_0 h_\mathrm{p} = 24.3\,\mathrm{mT}$ ($G = 1.2$). The amplitude has hereby been integrated in time over the forward traveling signal pulse in a time window of $\pm 4\,\mathrm{ns}$, i.e., for each pulse $n$ from $t = (t_0-4\,\mathrm{ns} + n \cdot \Delta t)$ to $t = (t_0+4\mathrm{ns} + n \cdot \Delta t)$. As can be seen from the linearity in the insets, the data show a clear exponential decay/growth, respectively. While the regime $G > 0$ is highly interesting for neuromorphic applications in general, since it provides easy access to nonlinearity, for the magnon adder, we chose the working point at $G \approx 0$. In this case, the energy inserted is just enough to compensate the losses and the current amplitude of the pulse within the resonator is preserved. Please note that due to the fact that the amplification is proportional to the amplitude, this compensation point holds for a small and a large amplitude spin-wave alike, as long as no nonlinearity sets in. In the following, we will fix the amplification field to $\mu_0 h_\mathrm{p} = 12.8\,\mathrm{mT}$, the field also used in Fig.~\ref{Fig2}~(b), to stay at $G \approx 0$.

\section{Demonstration of analog adding and subtracting}
\begin{figure}[h]
	  \begin{center}
    \scalebox{1}{\includegraphics[width=8 cm, clip]{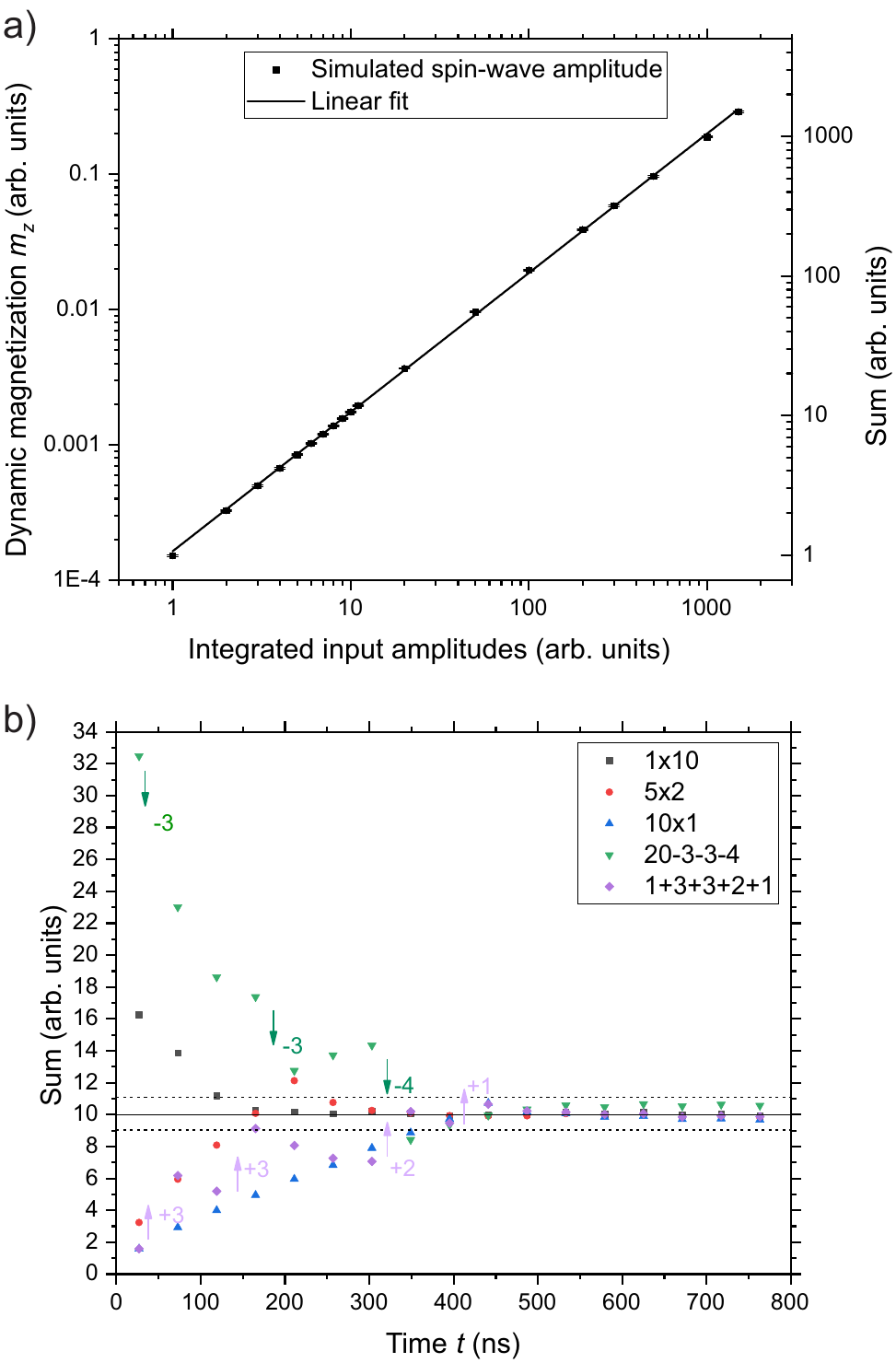}}
    \end{center}
	  \caption{\label{Fig4}a) Quasi steady-state amplitude in the resonator vs. input stimulus on a double-logarithmic scale. The straight line is a linear fit yielding a slope of $1.029\pm0.004$ confirming the linear relationship between the input amplitude and the sum in the resonator. b) Different combinations resulting in a sum of 10 within the resonator. Since the number of applied pulses as well as the time of their arrival is different in all cases, the sum of '10' is reached at different times for the different combinations.}
\end{figure}

For $G = 0$, the resonator losses are compensated. In this regime, a spin-wave pulse within it is cached as long as the amplification remains switched on and the compensated resonator can be used as a spin-wave adder. Since the spin-wave dynamics in the resonator are linear, the amplitude of the spin-wave pulse stored within the resonator corresponds to the sum $S$ over all incident pulses. This sum $S$ is given by $S = \sum_n A_\mathrm{p}(n) \cdot(-1)^{\phi(n)}$, where $A_\mathrm{p}(n)$ is the amplitude of the individual pulse $n$ and $\phi(n)$  represents its phase, being either $\phi(n) = 0$ for a phase-shift of $0$ or $2\pi$, and $\phi(n) = 1$ for a phase-shift of $\pi$. A phase-shifted pulse, thus, corresponds to a negative value and allows for a subtraction. For an input amplitude $A_\mathrm{p}(n)$ ranging from '0' to '100', individual pulses with the respective value of $A_\mathrm{p}$ can be applied at the input without significant nonlinear effects, corresponding to excitation field amplitudes ranging from $65\,\mu\mathrm{T}$ up to $6.5\,\mathrm{mT}$ in the input. Numbers $\leq 100$ can, therefore, be injected into the amplifier and will be summed over in an analog fashion. For larger excitation fields in the input, the spin-wave dynamics in the input become nonlinear, which distorts the summation. Nevertheless, within the amplifier, much larger numbers can be handled, since only a fraction of the input spin-wave is coupled into the resonator by the dipolar coupling. 
Figure \ref{Fig4}~(a) shows the amplitude of the spin-wave pulse within the resonator in the quasi steady-state, which corresponds to the sum $S$, as a function of the input amplitude on a double-logarithmic scale. As can be seen from the figure, the output is perfectly linearly proportional to the sum of the input amplitudes, no matter if an individual spin-wave pulse or a series of pulses is applied. This holds in the entire tested input range ranging from '1' up to at least '1500'. The latter corresponds to the sum of 15 pulses with an individual amplitude of '100', which are subsequently added in the resonator. The straight line is a linear fit yielding a slope of $1.029\pm0.004$, confirming the linear relationship between input and output. 

From the inputs exceeding '100', it can already be inferred from Fig.~\ref{Fig4}~(a) that the spin-wave packets in the resonator add up in a linear fashion. For instance, the output '200' corresponds to the sum of two input pulses of value '100', and so on. The key feature of the adder is that the device performs the summation purely analog, and the amplitude of the wave running back and forth in the resonator is directly proportional to the sum of the amplitudes of the input pulses. To illustrate this further, Fig.~\ref{Fig4}~(b) shows the amplitude in the resonator for several combinations adding up to 10: 1 $\times$ input '10', 5 $\times$ input '2', 10 $\times$ input '1', as well as the more involved pattern '1'+'3'+'0'+'3'+'0'+'0'+'0'+'2'+'1'. As mentioned above, the fact that spin-waves carry amplitude and phase can also be used to do a subtraction. This is also shown in Fig.~\ref{Fig4}~(b) by the combination '20'-'3'-'0'-'0'-'3'-'0'-'0'-'4'-'0'-'0', being equal to '10'. 

It should be noted that in the demonstrated regime of operation of the adder, the spin-wave dynamics stay linear. This allows to add up individual spin-wave pulses. The device already performs similar to a neuron in the brain: Incident pulses are converted into an amplitude information within the resonator and this amplitude is given by the integration over the incoming signals. In the presented adder, small pulses carrying the amplitude of the sum are constantly ejected into the output (cf. Fig.~\ref{Fig1}~(b)). Towards a neuromorphic application, the resonator could be designed in a way that its quality factor is a function of the spin-wave amplitude, for instance by a change of the dipolar coupling efficiency associated with a nonlinear change of wave-vector. This way, a nonlinearity can open up the resonator once the critical stimulus is overcome. In such a way, spin-wave axons that can be conveniently integrated into extended networks become feasible.

\section{Conclusion}
To conclude, by means of micromagnetic simulations, we have demonstrated a magnon adder, where the magnon amplitude adds and subtracts in an analog fashion. The spin-wave summation is performed in a resonator, whose losses are compensated by a parametric amplifier. This way, the amplitude is stabilized and constant in time if no mathematical operation is performed. The spin-wave signal in the resonator is directly proportional to the time-integrated amplitude of the incoming pulses. Hereby, the phase degree of freedom of the spin-waves allows to add spin-wave pulses in the case of constructive interference between the incoming spin-wave pulses. If a pulse is shifted by $\pi$, it will instead be subtracted. The presented device can perform as a magnon cache memory that can store an analog magnon sum on long time scales and, thus, constitutes the first step towards an all-magnonic neuron

\begin{acknowledgments}

The authors thank B. Hillebrands and A. Chumak for their support and valuable scientific discussion. They also gratefully acknowledge financial support by the DFG in the framework of the Research Unit TRR 173 \textit{Spin+X} (Projects B01), the Nachwuchsring of the TU Kaiserslautern as well as the ERC Starting Grant 678309 \textit{MagnonCircuits}.
\end{acknowledgments}

\end{document}